# High Pressure X-Ray Diffraction Study of $UMn_2Ge_2$

V. Siruguri[a,*], S.K. Paranjpe[b], P. Raj[c], A. Sathyamoorthy[c], J.-P. Itié[d], A. Polian[e]

[a] *Inter University Consortium for DAE Facilities - Mumbai Centre, R-5 Shed, Bhabha Atomic Research Centre, Mumbai 400 085, India*

[b] *Solid State Physics Division, Bhabha Atomic Research Centre, Mumbai 400 085, India*

[c] *Novel Materials & Structural Chemistry Division, Bhabha Atomic Research Centre, Mumbai 400 085, India*

[d] *LURE, Bât. 209D, Centre Universitaire Paris-Sud, 91405 Orsay, France*

[e] *Physique des Milieux Condensés, CNRS-UMR 7602, Université Pierre et Marie Curie, B 77, 4 Place Jussieu, 75252 Paris Cedex 05, France*

**Abstract**

Uranium manganese germanide, $UMn_2Ge_2$, crystallizes in body-centered tetragonal $ThCr_2Si_2$ structure with space group *I4/mmm*, $a$ = 3.993Å and $c$ = 10.809Å under ambient conditions. Energy dispersive X-ray diffraction was used to study the compression behaviour of $UMn_2Ge_2$ in a diamond anvil cell. The sample was studied up to static pressure of 26 GPa and a reversible structural phase transition was observed at a pressure of ~16.1 GPa. Unit cell parameters were determined up to 12.4 GPa and the calculated cell volumes were found to be well reproduced by a Murnaghan equation of state with $K_0$ = 73.5 GPa and $K'$ =

11.4. The structure of the high pressure phase above 16.0 GPa is quite complicated with very broad lines and could not be unambiguously determined with the available instrument resolution.

## 1  Introduction

$UMn_2Ge_2$ belongs to the general class of ternary intermetallic $UT_2X_2$ compounds, where T is a 3d transition metal and X is either Ge or Si. The crystal structure is tetragonal and of $ThCr_2Si_2$-type with the space group *I4/mmm*. These alloys have been extensively studied because of the rich variety in their magnetic properties[1-4]. The variation in the magnetic properties has been interpreted in terms of two parameters, namely, the inter- atomic distances between the U and T atoms, $d_{U\text{-}T}$, and 5*f*(U)-3*d*(T) hybridization due to change in 3*d* electron concentration[1]. It is generally seen that only the U sublattice orders magnetically and its nature and interaction strength varies from Pauli paramagnetic to a combination of different magnetic phases such as ferro and antiferromagnetic. $UMn_2Ge_2$ is an exception to this trend since both the Mn and U sublattices order ferromagnetically below 380 K and 100 K, respectively[3,5]. The distances between the U atoms are above the Hill limit which excludes the possibility of a direct overlap of the U 5*f* orbits, implying localized moments on the uranium site. However, hybridization with the overlapping bands of the 3*d* atoms can result in itinerant 5*f* states, depending on the distances between the U and 3*d* atoms. It was proposed to study the structural and magnetic properties of this system as a function of pressure, where one could observe the effects of the changed inter-atomic distance keeping the 3*d* electron concentration unaltered. Considering the fact that the magnetic behavior as a function of pressure is bound to evade a correct description as long as a complete knowledge of the structural behavior is lacking, a high pressure study of the structural properties of $UMn_2Ge_2$ becomes imperative. Hence, the present study aims to investigate the structural stability of $UMn_2Ge_2$ as a function of external pressure. In this paper, results of high pressure x-ray diffraction studies on $UMn_2Ge_2$ are presented.

## 2    Experimental

$UMn_2Ge_2$ was prepared by arc melting weighed amounts of U, Mn and Ge in stoichiometric proportions under Ar atmosphere. The buttons were flipped over and re-melted several times to ensure homogeneity after which they were annealed in vacuum. The annealed buttons were crushed into a fine powder and a laboratory x-ray diffraction pattern confirmed that the material was single phase with tetragonal structure and the lattice constants were in good agreement with the literature data[1,3]. High pressure diffraction studies at room temperature were undertaken using the energy dispersive x-ray diffraction (EDXRD) beamline at LURE synchrotron facility. A few micrograms of finely powdered sample were loaded into a membrane type diamond anvil cell (DAC) with silicone oil as the pressure transmitting medium. A few ruby spheres were included to measure the pressure by the ruby fluorescence method. A stainless steel gasket with a 200 $\mu$m hole was used. The EDXRD facility is situated at Station WDIS which uses the white beam from DW11A 5T wiggler beamline of DCI[6]. The beam was collimated down to 50 $\mu$m x 50 $\mu$m before the sample. After the DAC, a tungsten carbide slit 50mm long was used to define the diffraction angle. A Ge(Li) detector was set at a fixed scattering angle of 4.302° so as to maximize the known sample peaks in the energy range of 10 - 70 keV. A multi-channel analyzer was used to obtain the whole spectrum simultaneously and the high x-ray flux allowed short collection times per spectrum.

# 3    Results and Discussion

EDXRD spectra of UMn$_2$Ge$_2$ at some representative pressures up to 24 GPa are shown in Fig. 1. The spectra are normalized to the 13.6147 keV fluorescence peak of uranium at ambient pressure. It is observed that, at ambient pressure, all peaks other than the U fluorescence and escape peaks, could be indexed according to the tetragonal ThCr$_2$Si$_2$-type structure. With increasing pressure, the peaks shift to higher energies due to contraction of the lattice. Beyond 14.8 GPa, the fundamental peaks start broadening and at 16.1 GPa, there are clear indications of a phase transition taking place. The (112) reflection splits into two peaks and there are new peaks which appear at 21.5, 25.6, 26.8, 31.2, 35.5, 36.2, 38.2, and 39.9 keV (indicated by arrows in Fig. 1). As the pressure is increased beyond 16.1 GPa, (101) reflection progressively decreases and disappears while there is a further splitting of the peaks in the region of 36 keV and the new peak at 21.5 keV increases rapidly in intensity. The signatures of the original tetragonal phase are seen to persist up to 17.4 GPa. The disappearance of (101) peak is an important observation since this peak is crucial to the I4/mmm symmetry and its absence could mean that the body centering may not exist any more. These observations clearly show that there is a phase transition occurring around 16 GPa. The diffraction lines start overlapping and become very broad at 19.3 GPa pressure and at higher pressures, they continuously merge and split and the structure seems to stabilize only at 24.3 GPa. At this pressure, the lines are extremely broad with reduced intensity, except for the peak at 21.5 keV, and hence, attempts to index the diffraction pattern were futile. Also, the energy spectrum below 25 keV is dominated by U fluorescence lines, which makes it difficult to observe the low energy diffraction peaks from the sample. Therefore, within the resolution limits of the instrument, it is extremely difficult to arrive at the structure of high pressure phase. A remarkable observation is that when the pressure was released, the extra lines disappeared and the diffraction pattern reverted to the original tetragonal $ThCr_2Si_2$-type structure (Fig. 1) with slightly larger linewidths. Hence, this phase transition appears to be a reversible one.

Unit cell parameters up to 12.4 GPa pressure, where the tetragonal structure corresponding to the *I*4/*mmm* space group is retained, were refined from fitted peak positions using program PowdMult[7]. Variation of cell parameters with pressure is shown in Fig. 2 and it is observed that *a* and *c* smoothly decrease with increasing pressure. The compression along *c* is slightly larger (3.7%) compared to compression along *a* (2.5%). The pressure-volume data shown in Fig. 3 were first fitted with a Murnaghan equation of state[8] (EOS) given by:

$$P = K_0 / K' \left[ (V/V_0)^{-K'} - 1 \right]$$

where $K_0$ is the isothermal bulk modulus and $K'$ its pressure derivative. The reference volume, $V_0$, was 172.17Å$^3$, which is the volume at ambient pressure. Initially, the value of K' was constrained to 4 and the least squares fit gave a value of 100 GPa for the bulk modulus. The result of this fit is shown as dashed line in Fig. 3. Then, the value of K' was allowed to vary, which gave a value of 73.5 GPa for $K_0$ and 11.4 for K'. The result of least squares fit is shown by the full curve in Fig. 3. It is clear from the two curves that the value of K' cannot be constrained to 4. The data were also fitted with a third-order Birch-Murnaghan EOS [8,9] and the results were comparable to that of Eq. (1). It must be noted here that the pressure transmitting medium, silicone oil, solidifies in the pressure range 8 to 10 GPa. Including data in this pressure range could lead to artificial modifications of the bulk modulus. However, there is no such indication in the present data. Fig. 4 shows the variation of the *c/a* ratio with pressure. The data shows, within the scatter, a decreasing trend in the *c/a* ratio up to 12.4 GPa (dashed line). There is a sudden jump in the ratio for higher pressures, clearly indicating the onset of a structural phase transition.

The $K_0$ value of 73.5 GPa obtained from Eq. (1) is much smaller than the value of 215 GPa found for the compound, $URu_2Si_2$ [10] (which has the same space group and comparable lattice parameter values) and of 168 GPa found for the compound, $YbCu_2Si2$ [11]. This result shows that the present compound is much "softer" than $URu_2Si_2$, where Ru is a 4$d$ transition element, and $YbCu_2Si_2$, where Yb is a rare-earth element. The K' value of 11.4, which is on the higher side, is not unusual for uranium compounds. It has been observed earlier that $USi_3$ and $USn_3$ [12] and URhAl [13] have comparable values for the bulk modulus and its pressure derivative. Higher values of bulk modulus in uranium compounds have been associated with increased itinerancy of the 5$f$ electrons [14]. Therefore, the lower value of bulk modulus in the present compound could imply that the degree of itinerancy of the 5$f$ states is much less as compared to other uranium compounds. It is well known that for $UMn_2Ge_2$ even though both 5$f$ and 3$d$ bands are near the Fermi level, the distance between U and Mn ($d_{\{U-T\}}$) is largest among its isostructural compounds at 6.35 a.u. This practically nullifies the hybridization effects so that both bands can give rise to separate existence of magnetic moments [1], [15]. However, with increasing pressure it is likely that the hybridization effects start dominating and lead to decrease of magnetic moments. It has been observed that α-Mn has $K_0$ and K' values of 158 GPa and 4.6, respectively, over a pressure range from ambient to 165 GPa [16]. It was also observed that $K_0$ value of α-Mn systematically increases depending on the increase of pressure range to be fitted, indicating decrease of magnetic moments under pressure. Hence, it is probable that the observed phase transition around 16 GPa is associated with a pressure induced delocalization of the 5$f$ states. High pressure XRD measurements with better resolution are necessary to solve the structure of the high pressure phase and determine its EOS, which could lead to more information on the 5$f$ electronic structure of the high pressure phase.

## 4 Conclusions

In conclusion, it is observed from the compression behavior of $UMn_2Ge_2$ that the *I*4*/mmm* structure remains stable up to 12.4 GPa. The value of the bulk modulus is found to be much lower than those obtained for similar compounds, suggesting that the hybridization of the U 5*f* states may not be sufficiently strong and that these states may effectively remain localized to a large extent. Around 16 GPa, the *I*4*/mmm* structure becomes unstable and there is a phase transition as evidenced by the splitting of the fundamental reflections and appearance of new peaks. This transition is reversible since the original spectrum is regained on release of pressure. The structure of the high pressure phase is rather complicated and further measurements at higher resolutions are necessary to arrive at the correct structure and the corresponding EOS.

## 5 Acknowledgments

Financial support by Department of Science and Technology and Indo French Center for Promotion of Advanced Research, India for carrying out the above work is gratefully acknowledged. One of us (VS) wishes to thank Drs. B.A. Dasannacharya and P.S. Goyal for their encouragement and interest in this project.

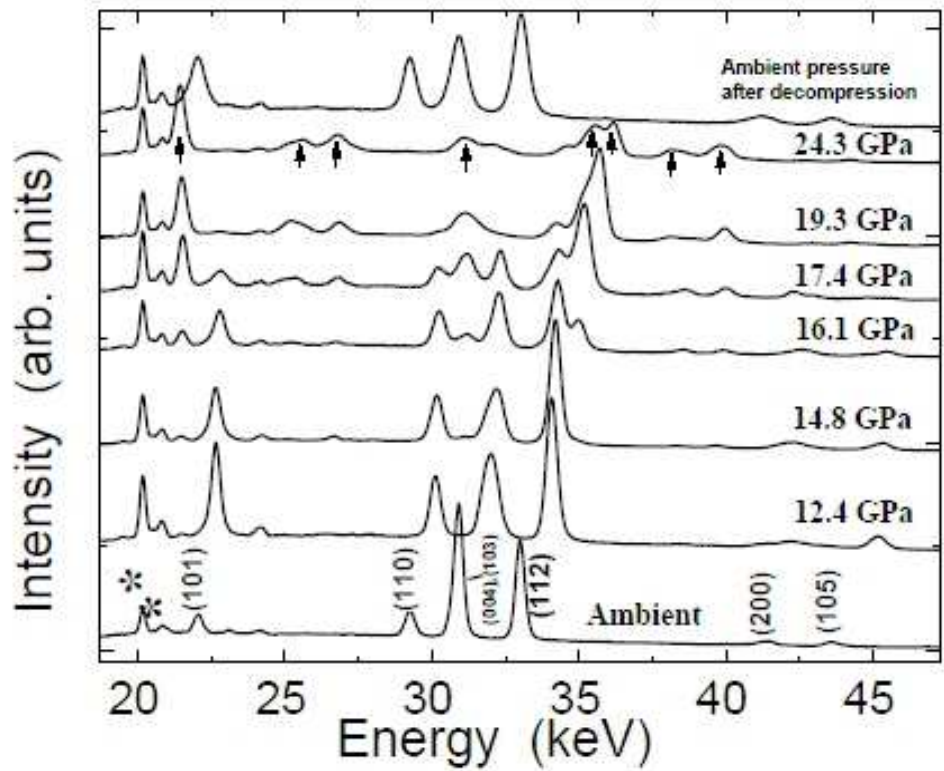

Fig. 1. EDXRD spectra of $UMn_2Ge_2$ at different pressures. Stars and arrows indicate U fluorescence peaks and appearance of new lines, respectively.

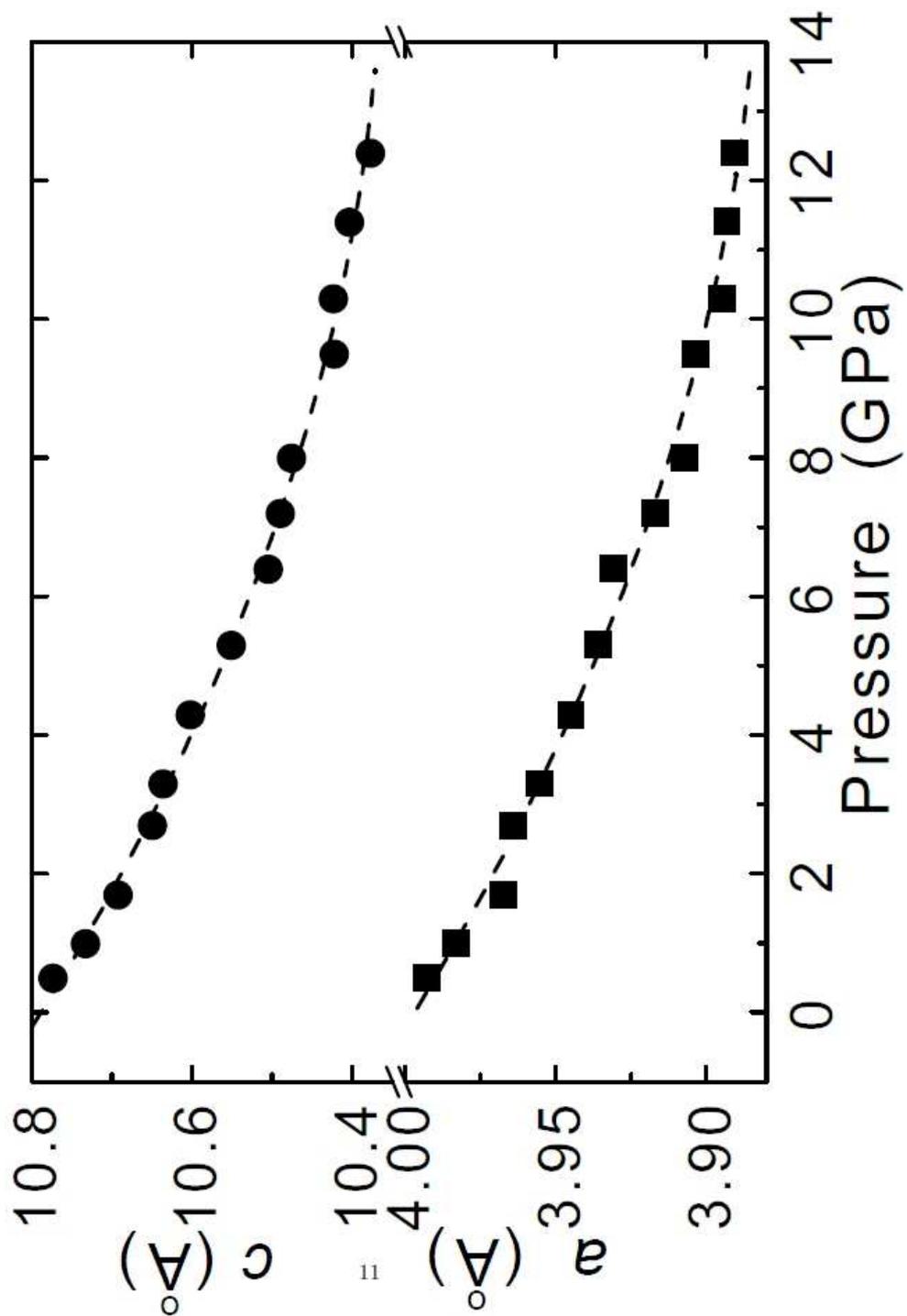

Fig. 2. Effect of cell parameters *a* and *b* up to 12.4 GPa. The dashed lines are guides for the eye.

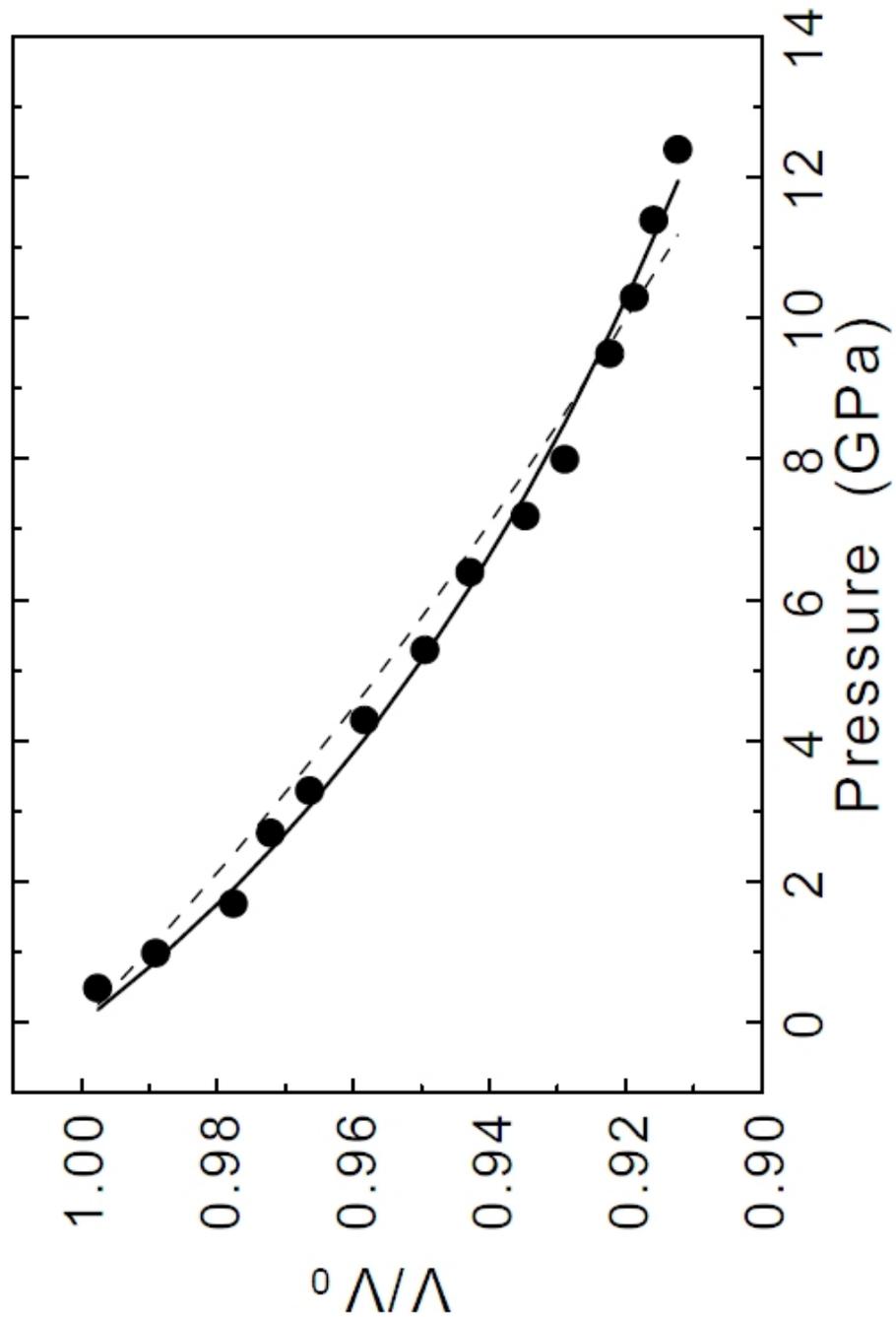

Fig. 3. Relative volume versus pressure for $UMn_2Ge_2$. The dashed curve through the data points is a fit to the Murnaghan EOS (Eq. 1) with K' = 4 (constrained) and the continuous curve is a fit to the same EOS without any constraints.

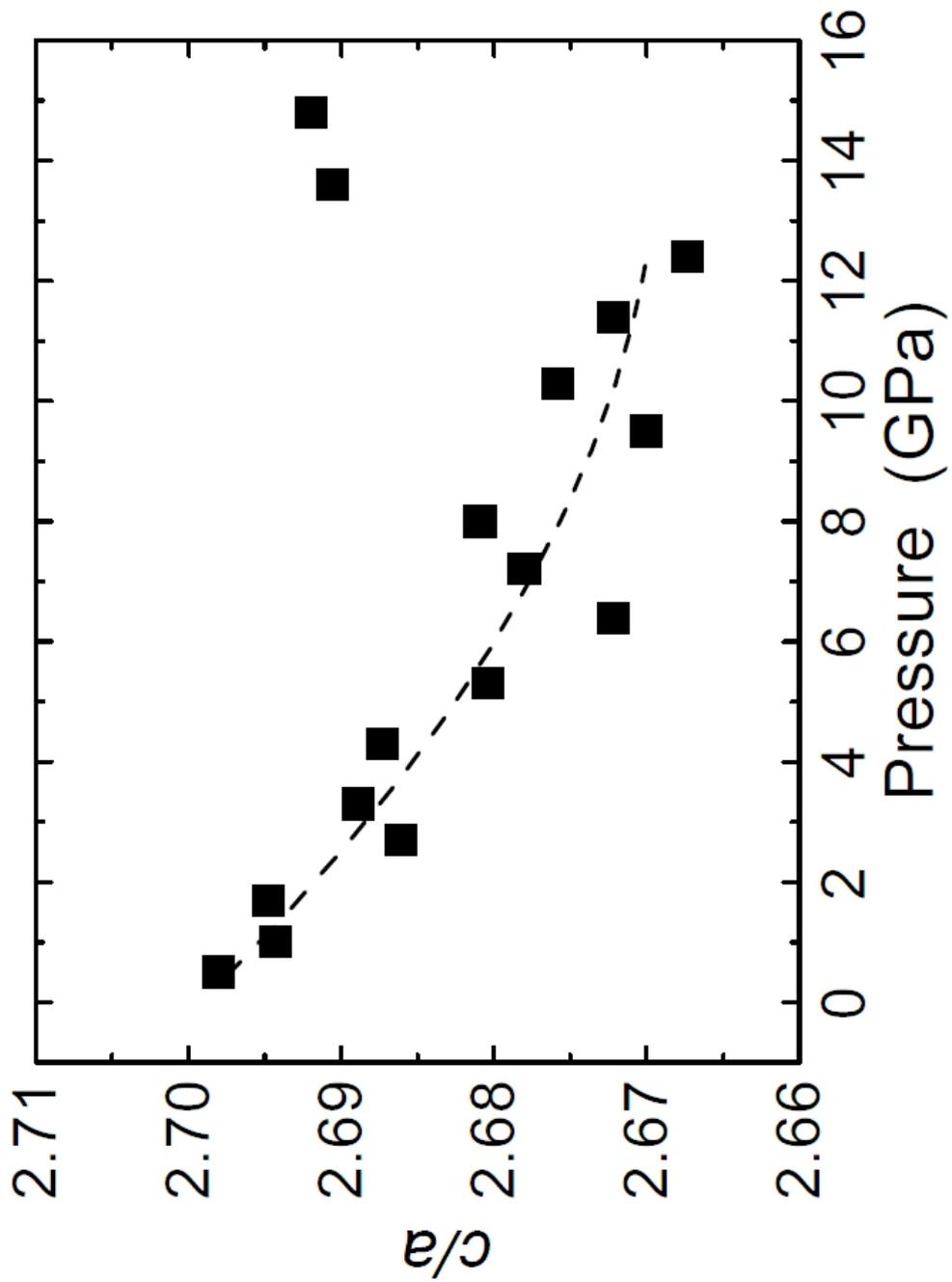

Fig. 4. *c/a* ratio versus pressure for $UMn_2Ge_2$. The dashed curve through the data points is a guide to the eye.